\documentclass[11pt,a4paper]{article} 
\pdfoutput=1
\usepackage{amsmath}
\usepackage{enumerate}
\usepackage{amssymb}
\usepackage{graphicx}
\usepackage{multirow}
\usepackage{xcolor}
\usepackage{bm}
\usepackage[normalem]{ulem}
\usepackage[autostyle]{csquotes} 

\usepackage{booktabs}
\usepackage{color}
\usepackage{jcappub}

\def\m@th{\mathsurround=0pt }
\def\eqalign#1{\null\,\vcenter{\openup1\jot \m@th
 \ialign{\strut\hfil$\displaystyle{##}$&$\displaystyle{{}##}$\hfil
 \crcr#1\crcr}}\,}

\title{Neutrino Masses and Mass Hierarchy: Evidence for the Normal Hierarchy}

\author[1,2]{Raul Jimenez,}
\author[3,4]{Carlos Pena-Garay,}
\author[1]{Kathleen Short,}
\author[5]{Fergus Simpson}
\author[1,2,\#]{and Licia Verde}

\affiliation[1]{ICC, University of Barcelona, IEEC-UB, Mart\' i i Franqu\` es, 1, E08028
Barcelona, Spain}
\affiliation[2]{ICREA, Pg. Lluis Companys 23, Barcelona, 08010, Spain} 
\affiliation[3]{Laboratorio Subterr\'aneo de Canfranc, 22880 - Estaci\'on de Canfranc, Huesca, Spain.}
\affiliation[4]{I2SysBio, CSIC-University of Valencia, 46071 - Valencia, Spain.}
\affiliation[5]{SecondMind, 72 Hills Road, Cambridge, CB2-1LA, UK.}
\affiliation[\#]{corresponding author}

\emailAdd{raul.jimenez@icc.ub.edu; cpenya@lsc-canfranc.es;katie.short@icc.ub.edu;fergus@secondmind.ai; liciaverde@icc.ub.edu}

\abstract{The latest cosmological constraints on the sum of neutrino masses, in combination with the latest laboratory measurements on oscillations, provide ``decisive" Bayesian evidence for the normal neutrino mass hierarchy.  We show  that this result  holds across very different prior alternatives  by exploring two extremes on the range of prior choices.  In fact,  while the specific numerical value for the Evidence depends on the  choice of prior,  the Bayesian odds  remain  greater than $140:1$  across very different prior choices.   For Majorana neutrinos this has important implications for the upper limit of the neutrino-less double beta decay half life and thus for the technology and resources needed for future double beta decay experiments.} 

\begin{document}
\maketitle
\flushbottom

\section{Introduction} 

Solar, atmospheric, accelerator and reactor neutrino experiments have observed neutrino flavor conversion driven by neutrino masses, but the observable effects of neutrino oscillations are sensitive only to mass-squared splittings among eigenstates and thus can only constrain the neutrino mass square differences. Matter effects have allowed the sign of the small mass 
splitting to be determined, leaving two possible mass orderings: normal hierarchy (NH) and inverted hierarchy (IH). Determining the precise values of the three mass states, and the mass ordering (i.e., the hierarchy) or, equivalently, the precise determination of the sum of the masses (or the absolute mass scale) and the hierarchy, remains an open challenge. 

The absolute neutrino mass scale can be probed using beta decay; the best current bound is provided by the KATRIN experiment with an upper limit on the beta decay effective neutrino mass of $m_{\nu,\beta}< 0.8$ eV, i.e., the sum of neutrinos masses $\Sigma < 2.4$ eV~\cite{katrin21} at 90\% confidence, which improves dramatically upon the previous limit of $\Sigma < 6.9$ eV at 95\% confidence~\cite{mainz}. Yet, despite these spectacular improvements, this bound is still about an order of magnitude less stringent than state-of-the-art cosmology constraints.

This is where cosmological observations become of relevance (see e.g,~\cite{Bond:1980ha,Hu:1997mj,1998PhRvL..80.5255H,lesgourguespastor,JimenezKitching,Wagner:2012sw} and references therein). Cosmological surveys can provide crucial information on the absolute masses of neutrinos, as massive neutrinos influence both the expansion history and growth of structure in the Universe. Over the past five years the cosmological limits on the sum of neutrino masses, $\Sigma$, have become increasingly tighter, and are getting tantalizingly close to the lower limit for the sum of the masses allowed by the inverted hierarchy $\Sigma_{{\rm IH},low}$ = 0.0982 ± 0.0010 eV (68\% C.L.) \cite{eBOSSresults,1911.09073}, implying the volume of parameter space available for the IH is becoming heavily restricted. Here we revisit the question \cite{Fergus}: ``Given the current knowledge about mass-squared splittings and cosmological constraints on the total mass, can anything be said about the neutrino mass hierarchy?".

Whether the electron neutrino is mainly composed by the lightest mass state (as in the NH) or not (IH) has important implications on the theoretical upper limit of the neutrino-less double beta decay half life~\cite{Murayama:2003ci}, in the case that neutrinos are Majorana fermions and obtain the Majorana mass through the three light neutrinos mechanism. Therefore, the fraction of the neutrino parameter space to be explored by future ton-scale double beta decay experiments~\cite{Giuliani:2019uno} is greatly influenced by the type of neutrino mass hierarchy, with implications for future technologies and resources needed that may lead to the discovery.

In cosmology we cannot perform experiments; we only observe the sky. As such, we always test a model of the observations that we gather from the sky with our detectors.
Given this constraint, the most natural choice to make inference in cosmology and astronomy is Bayesian statistics: we want to infer the parameters of the model given some data. This takes the following, extremely well known, mathematical form
\begin{equation}
P(\alpha| D,M) = \frac{P(D|\alpha, M) P(\alpha|M)}{P(D|M)},
\label{eq:bayes}
\end{equation}
where $D$ stands for data (observations), $M$ for model (hypothesis) and $\alpha$ denotes the parameters of the model; $P(\alpha| D,M)$ is the posterior, used for parameter inference,  $P(D|\alpha,M)$ is the likelihood which is usually provided along with the data. $P(\alpha|M)$ is the prior, the so-called prior knowledge, which has provoked abundant literature for centuries on its choice and value and an on-going debate between Bayesian and frequentist approaches to probability. All Bayesian inference depends on the choice of the prior, which is \emph{always} a subjective choice. It is true that for parameter inference, in the limit of precision measurements -- i.e., if the likelihood is very localized -- the likelihood can overcome (most reasonable choices of) the prior.

The question we set out to address, however, falls under ``model selection" or ``model comparison"  and therefore it is useful to employ Bayesian evidence methods~\cite{Cox}, which have gained attention in cosmology over the past couple of decades~\cite{Jaffe:1996, Trotta2,LiddleEvidence04}. The evidence can be written as a function of the likelihood and the prior, being the integral of the likelihood over the (full range of the) prior integrated over the the values of the parameters $\alpha$:
\begin{equation}
    P(D|M)=\int P(D|\alpha,M)P(\alpha|M)d\alpha\,.
    \label{eq:Evidence}
\end{equation}
The evidence quantifies the probability of obtaining the data given the full model, not just a specific set of parameters values.
What we really want to infer is the probability of a particular model given the data $ P(M|D) = P(D|M) P(M)/P(D)$; 
when considering two models, each of them with the same a priori probability $P(M)$ for the same data, model comparison can be done by taking the evidence ratio for the two models. 
It is important to note that the evidence  and hence also the evidence ratio will \emph{always} depend on the choice of the prior $P(\alpha|M)$, even for very localized likelihoods. 

In light of this limitation, it is important to fully understand the motivation for and implications of a given prior choice, the impact of the prior choice on the inferred result or, alternatively, how to choose the prior depending on the context or the question at hand.

\blockquote{Historically, Bayesian analysis has been accompanied by methods to work out the ``right" prior for a problem, for example, the principles of insufficient reason and maximum entropy. The modern Bayesian, however, does not take a fundamentalist attitude to assigning the ``right" priors - many different priors can be tried; each particular prior corresponds to a different hypothesis about the way the world is. We can compare these alternative hypotheses in the light of the data by evaluating the evidence.  The answer to this question can be reached by performing Bayesian model comparison. - DJC MacKay~\cite{mackay2003information}}.

Choosing an appropriate prior in the context of inferring the odds of one mass hierarchy over the other has been discussed at length in the literature~\cite{objectivebayesian, Fergus, referee_request, Blennow:2013kga, HannestadSchwetz, Vagnozzi17, LongRaveri, ChoudhuryHannestad20}. 
The initial publication ref.~\cite{Fergus} elicited strong reactions and stimulated a vivid discussion on the hierarchical nature of neutrinos. Many different sets of priors have been explored yielding a wide range of conclusions.
The disparities in the findings of these works further highlights the fact that the evidence {\it always} depends on the prior choice. Yet very few have followed MacKay's advice regarding how the prior choice encompasses our \emph{a priori} views about the way the world is -- in this case which set of prior is supported by our knowledge of e.g., physics and how that informs model selection.

For example, ref.~\cite{LongRaveri} argues that the prior choice should be motivated by fundamental physical principles and chooses to specify the prior directly at the level of the neutrino mass matrix, obtaining odds of the order of $\sim$100:1 for the NH, using the constraints from 2018. Ref.~\cite{Fergus} on the other hand, tries to be more generic and argues that the prior should describe our state of belief before we have constraints, i.e.~before the measurement of neutrino oscillations and thus before having information about the mass-squared splittings, when each of the neutrino masses has an uncertainty that spans many orders of magnitude. In this case, ref. \cite{Fergus} argues, a logarithmic prior on each of the masses naturally incorporates this uncertainty. More precisely, rather than using a single pre-determined prior, ref.~\cite{Fergus} adopts a family of priors described by hyper-parameters, yielding a hierarchical prior (see appendix~\ref{sec:appendixBHM}) and obtaining odds of 40:1 for the NH. Such a choice of prior reflects the fact the three masses are indistinguishable before the data arrives, and specifies the prior on each mass before any ordering takes place. 
This prior is well known in the particle physics community. It is based on the construction of the Standard Model mass mechanisms, which generally assume a common mechanism (usually at a high energy scale) for the origin of the three neutrino masses. In perspective, this prior is behind the rise of neutrino oscillation experiments.
The additional ingredient of adopting a hierarchical prior (see appendix \ref{sec:appendixBHM}) means that effectively rather than a given prior, a family of priors- described by  hyperparameters- is adopted,  and  the hyperparameters values are marginalized over.

A completely different approach to the same question, from the alternative school of thought that is ``Objective Bayesianism", is taken in ref.~\cite{objectivebayesian}. The idea is to construct a prior that is as uninformative as possible from a mathematical point of view. To do so the likelihood is used and in particular the likelihood provided by oscillation data. The adopted prior, for precision experiments where the likelihood achieves asymptotic normality, uses as information measure the Fisher information. This choice of prior depends on the experimental set up. In particular, in the case of neutrino hierarchy, it does not enclose any \emph{a priori} information about the fact that e.g., we believe there to be a single mechanism that gives rise to the three neutrino masses and that their masses can span many orders of magnitude (as in in ref.~\cite{Fergus}).
Still it is important to note that this prior  choice
 reflects the fact the three masses are indistinguishable before the data arrives, hence does not favor one hierarchy over another.\footnote{ Note that  conversely, it is possible to set up a prior that does not distinguish between the hierarchies yet does not satisfy exchangeability (the three masses are not drawn  from a common prior, see for example the priors adopted by \cite{GariazzoMena19, Hergt, Mahony} and other references therein). This choice   implies that  the three masses do not share a common origin  and therefore  that
 there are different physical  mechanisms that  give neutrinos the different masses. We do not consider this case. }
This \emph{objective} prior does need to be updated for every different experiment that is performed. For the experiments considered in ref.~\cite{objectivebayesian} this approach yielded odds of 5:1 for the NH.

Also, as MacKay points out, the very same framework that we use to compare the normal hierarchy to the inverted hierarchy  -- the Bayesian evidence ratio -- can also be used to identify which set of priors is favoured by the data. This can be explicitly explored in the hierarchical prior case, where the distribution of the hyper-parameters quantifies exactly this (see appendix \ref{sec:appendixBHM}). 

In this paper we revisit this issue, which has remained somewhat  dormant since 2018, and follow MacKay in considering two generic type of priors: the one used in ref.~\cite{Fergus} (the logarithmic-hierarchical prior, which we abbreviate by `SJPV' from the author's initials) and the one used in ref.~\cite{objectivebayesian} (the objective Bayesian or `HS'). 
We compute the evidence under each of these approaches given the latest data from oscillation experiments and cosmological observations, and reflect on how it changes. We then speculate on the implications for double beta decay experiments. While other prior choices have been considered in preprints and in the literature,  they are not of interest here because they are either not physically motivated,  or adopt different prior distributions for the different mass eigenstates or would return odds in between the HS and the SJPV cases.

\section{Data and methods}
\label{sec:datamethods}
Following standard practice we denote the three neutrino mass eigenstates as $m_1$, $m_2$, and $m_3$ such that there are two independent neutrino mass-squared differences and two possible hierarchies only. Here we adopt the convention where  $m_1<m_2$ and $m_1, m_2$  refers to the smaller mass-squared difference. 
Hence in the so-called normal mass hierarchy (NH), $m_1$, $m_2$, and $m_3$ are defined in ascending order such that $m_1 \leq m_2  \leq  m_3$.  In the inverted hierarchy (IH) the ordering is instead $m_3  \leq  m_1  \leq  m_2$. The hierarchy is given by the sign of the square mass splitting involving $m_3$. When needed, we might refer to the three masses as $m_{\rm L}$, $m_{\rm M}$ and $m_{\rm H}$ (for 'low', 'medium' and 'high' respectively). In the NH then $m_{\rm L} \lesssim m_{\rm M}\ll m_{\rm H}$, while in the IH instead $m_{\rm L}\ll m_{\rm M} \lesssim m_{\rm H}$.

We will use the following constraints on the squared mass splitting, derived from a global fit to observations of neutrino oscillations using NuFIT v5.1 (2021)\footnote{The original reference reports slightly asymmetric error bars:$^{+0.21}_{-0.20}$ for $\Delta m_{21}^2$ and $^{+0.026}_{-0.028}$ for $\Delta m_{32}^2$ IH,  which we have symmetrized here as Gaussianity is a key assumption of ref.~\cite{HeavensEvidence}.}~\cite{gonzalez-garciaglobal21}

\begin{equation}
\begin{gathered}
\label{eq:splittings}
\Delta m_{21}^2=m_2^2 - m_1^2 = 7.42  \, (\pm 0.21) \times 10^{-5} \mathrm{eV}^2 \,\,\, (68.4\% {\rm CL})\, \\[5pt]
 \Delta m_{3\ell}^2 = m_3^2 - m_\ell^2 = 
\begin{cases}
  \phantom{-}   2.510  \, (\pm 0.027) \times 10^{-3} \mathrm{eV}^2   & \text{(NH) }  \\
     -2.490 \, (\pm 0.027) \times 10^{-3} \mathrm{eV}^2        & \text{(IH)}
\end{cases}
\,\,\, (68.4\% {\rm CL})\,,
\end{gathered}
\end{equation}
where $m_\ell$ denotes  $m_1$ and $m_2$ for the normal (NH) and inverted (IH) hierarchies respectively, and we approximate the uncertainty distribution as a Gaussian. In particular we have symmetrized the error bars for simplicity as this has a completely negligible effect on the final results.  Compared to refs.~\cite{Fergus,objectivebayesian} the error bars have increased by $\sim$10\% on $\Delta m_{21}^2$  but decreased by $\sim$30\% on $\Delta m_{3\ell}^2$. This provides (a Gaussian approximation to) the likelihood $P(D|\alpha,M)$ where $\alpha$ denotes the mass-squared splittings of eq.~\eqref{eq:splittings}.

It is important to note that, as discussed extensively in ref.~\cite{gonzalez-garciaglobal21}, the three flavour oscillation parameters from the fit to global data yields the best fit for the normal ordering, with the solution for the inverted ordering having a worse fit by $\Delta \chi^2=7 \,(2.6)$ when including (not including) the Super-Kamiokande atmospheric data (SK-atm). Moreover, there is a small tension between T2K and NOvA data, where the normal ordering option preferred by NOvA is in tension with the combination of T2K and reactor neutrino data. Although the inclusion of Super-Kamiokande data in the analysis of ref.~\cite{gonzalez-garciaglobal21} is sub-optimal, we see no reason to exclude this data set, and hence our baseline results will always include SK-atm. For our purposes the mild tension with NOvA data and the sub-optimal inclusion of SK-atm primarily affect the $\Delta \chi^2$ values, which generally underestimate the power of disentangling the neutrino parameter space when the tension is resolved.

\begin{figure}[tbp]
\centering
\includegraphics{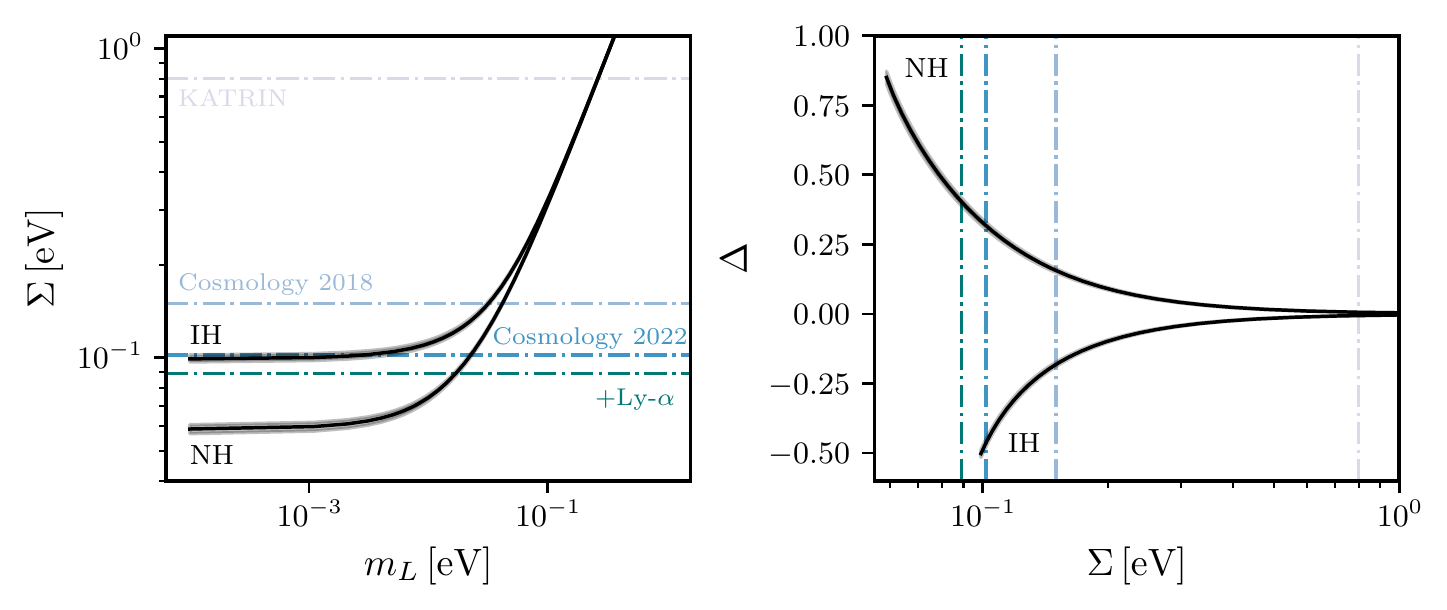}
\caption{Visualization of the neutrino oscillation experiment constraints (eq.~\eqref{eq:splittings}) and $\Sigma_{\beta}$ KATRIN sensitivity limit and $\Sigma_{\rm cosmo}$ constraints (95\% C.L.) in the $m_{\rm L}$-$\Sigma$ plane and $\Delta$-$\Sigma$ plane. The grey bands represent $5\sigma$ uncertainties on the oscillations measurements.}
\label{fig:oscillhierarchy}
\end{figure}

The oscillation constraints of eq.~\eqref{eq:splittings} can be visualized in figure~\ref{fig:oscillhierarchy} where we show the allowed region in the $m_{\rm L}$-$\Sigma$ plane, and in the $\Delta$-$\Sigma$ plane where $\Delta$ is defined as the ratio of the largest mass splitting to the total mass~\cite{JimenezKitching},
 \begin{equation}
 \Delta_{\rm NH}\equiv(m_{\rm H}-m_{\rm L})/\Sigma\,\,\,{\rm and}\,\,\, \Delta_{\rm IH}\equiv(m_{\rm L}-m_{\rm H})/\Sigma.
 \end{equation}

The bound on the sum of the masses $\Sigma=m_1+m_2+m_3$ derived from beta decay has significantly improved recently from $\Sigma <6.9$ eV (95\% C.L.) from the Mainz experiment~\cite{mainz} to $\Sigma <3.3$ eV (90\% C.L.) from the first KATRIN campaign~\cite{katrin2019} and even further to $\Sigma <2.4$ eV (90\% C.L.) from the combination of the first and second KATRIN campaigns~\cite{katrin21}. The estimated sensitivity of KATRIN is $\Sigma <0.8$ eV (95\% C.L.)
Meanwhile, over the past two decades cosmological surveys have yielded increasingly stronger constraints on the sum of the masses: 
$\Sigma_{\rm cosmo}<1.8$ eV in 2002~\cite{Elgaroy:2002bi}, $\Sigma_{\rm cosmo}<0.44$ in 2012~\cite{WMAP:2012nax},
$\Sigma_{\rm cosmo}<0.25$ eV~\cite{Planck2013cosmo} in 2013,  $\Sigma_{\rm cosmo}<0.18$ eV~\cite{Planck2015cosmo} in 2015 and $\Sigma_{\rm cosmo}<0.13$ eV~\cite{2016Cuesta} in 2016 (all quoted at 95\% C.L.). The latest constraints provided by the eBOSS collaboration from the joint analysis of cosmic microwave background and large-scale structure data obtain an upper limit of $\Sigma_{\rm cosmo}=0.102$ eV (95\% C.L.) for the combination CMB+BAO+RSD (Planck data, with baryon acoustic oscillations and redshift space distortions from e-BOSS) and $\Sigma_{\rm cosmo}=0.099 $ eV (95\% C.L.) for the combination CMB+BAO+RSD+SNe (including supernovae type 1A data)~\cite{eBOSSresults}. Finally in combination with the Lyman-$\alpha$ forest 1D flux power spectrum, ref.~\cite{1911.09073} obtains an even stronger bound of $\Sigma_{\rm cosmo}<0.089$ eV (95\% C.L.). We illustrate a few of these cosmological constraints in figure~\ref{fig:oscillhierarchy}.

In what follows, as is commonly done, we assume Gaussianity in all the reported constraints. The upper limit on $\Sigma_{\rm cosmo}$ can in principle be interpreted in two ways: either as a (one-sided) Gaussian distribution centered at $\Sigma=0$ or as a Gaussian distribution extrapolated to $\Sigma<0$ and centered at the (interpolated or extrapolated) maximum of the posterior (even if that happens to be at $\Sigma<0$) and truncated at $\Sigma=0$. Since the data show no indication of a detection of $\Sigma$ and negative masses are unphysical, we adopt the first interpretation. However, we explore the sensitivity of the results to this assumption for a few select cases. 

The priors and evidences for logarithmic hierarchical SJPV and objective Bayesian HS are computed following refs.~\cite{Fergus} and \cite{HeavensEvidence} respectively, using the updated constraints. In the case of HS, the interpretation of the mass splittings and the naming conventions of the individual masses are slightly different from that of the original paper to make them consistent with the treatment of SJPV and ref.~\cite{gonzalez-garciaglobal21}. Details are reported in appendices~\ref{sec:appendixBHM} and \ref{sec:appendixOBP}.

Different sources are responsible for different contributions to the final evidence ratio. We denote the final evidence ratio for the NH over the IH by $K_{\rm HS/SJPV}$ with subscript indicating the choice of prior. Partial results towards the full $K$, obtained including only some of the different contributions, are denoted by $\kappa_i$ (where the subscript $i$ specifies the  different contributions considered or excluded). In particular, the contribution to $K$ due to the $\Delta \chi^2$ of the global fit is a multiplicative factor: $\kappa_{\Delta \chi^2}=\exp(\Delta\chi^2/2)$. For completeness and a more transparent comparison with previous work, we also report the partial results excluding the $\kappa_{\Delta \chi^2}$ contribution. However, we stress here that we see no reason to exclude it. 

\begin{figure}[h!]
\centering
\includegraphics[width=.58\linewidth]{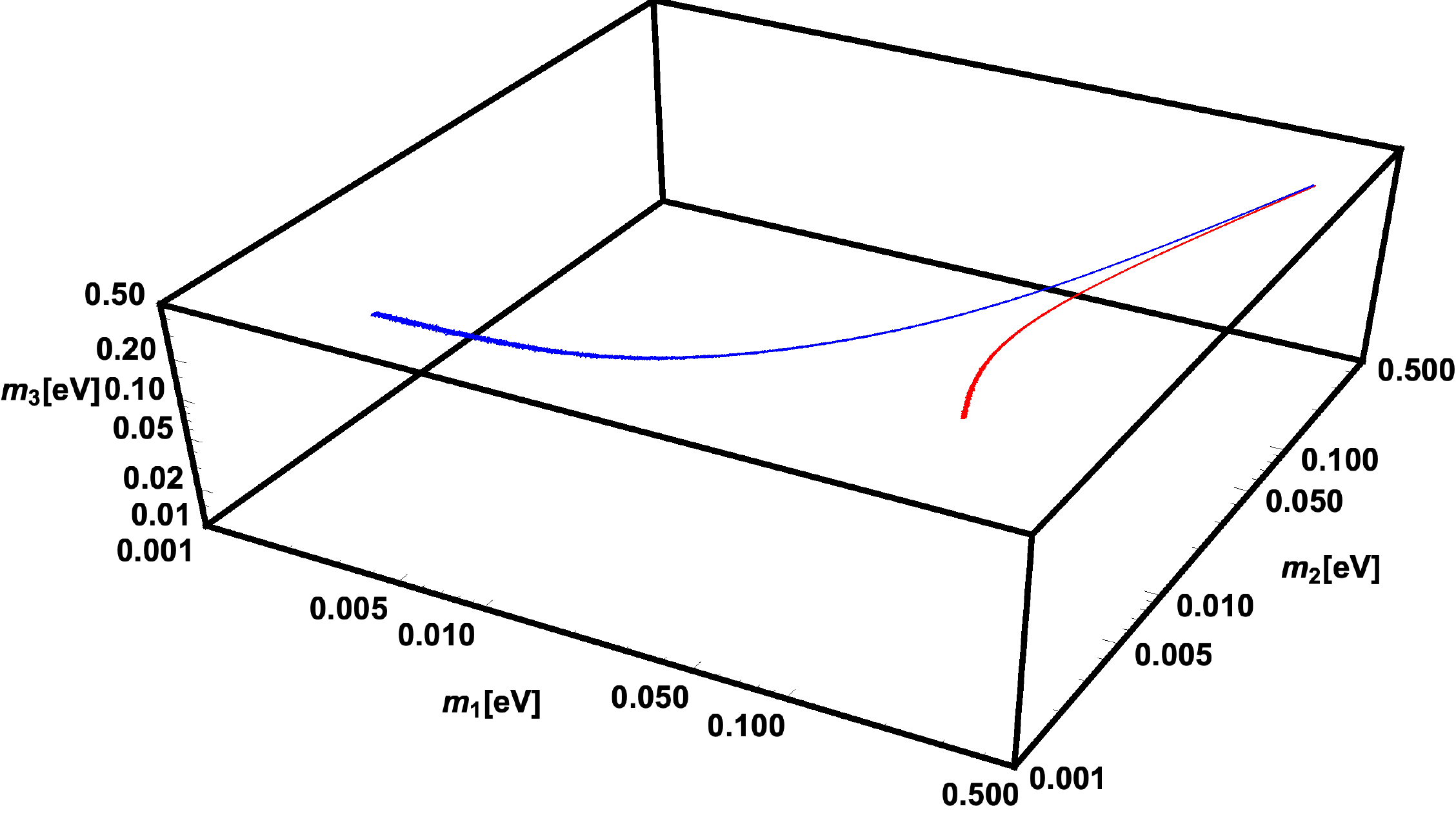}
\includegraphics[width=.58\linewidth]{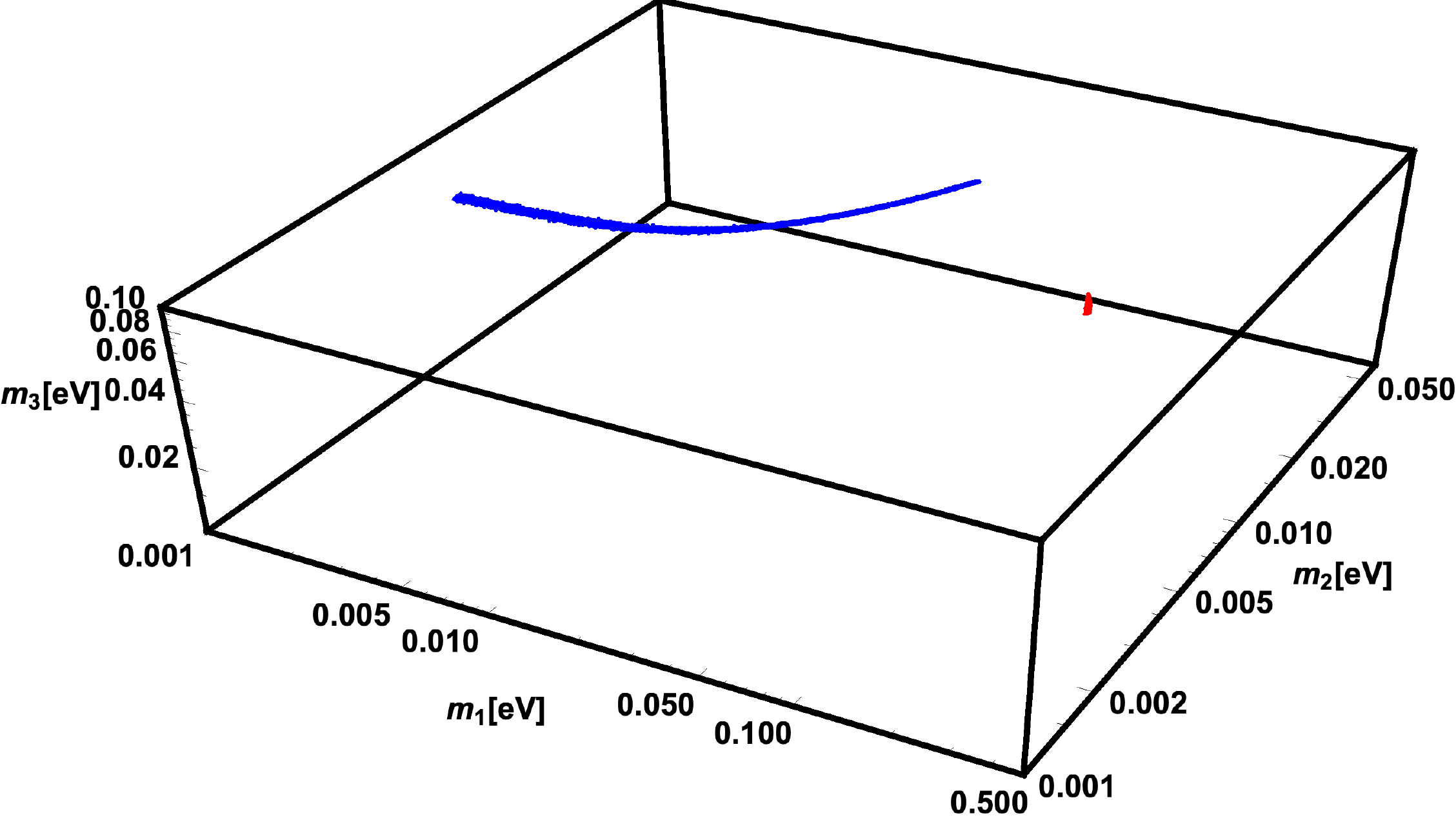}
\includegraphics[width=.5\linewidth]{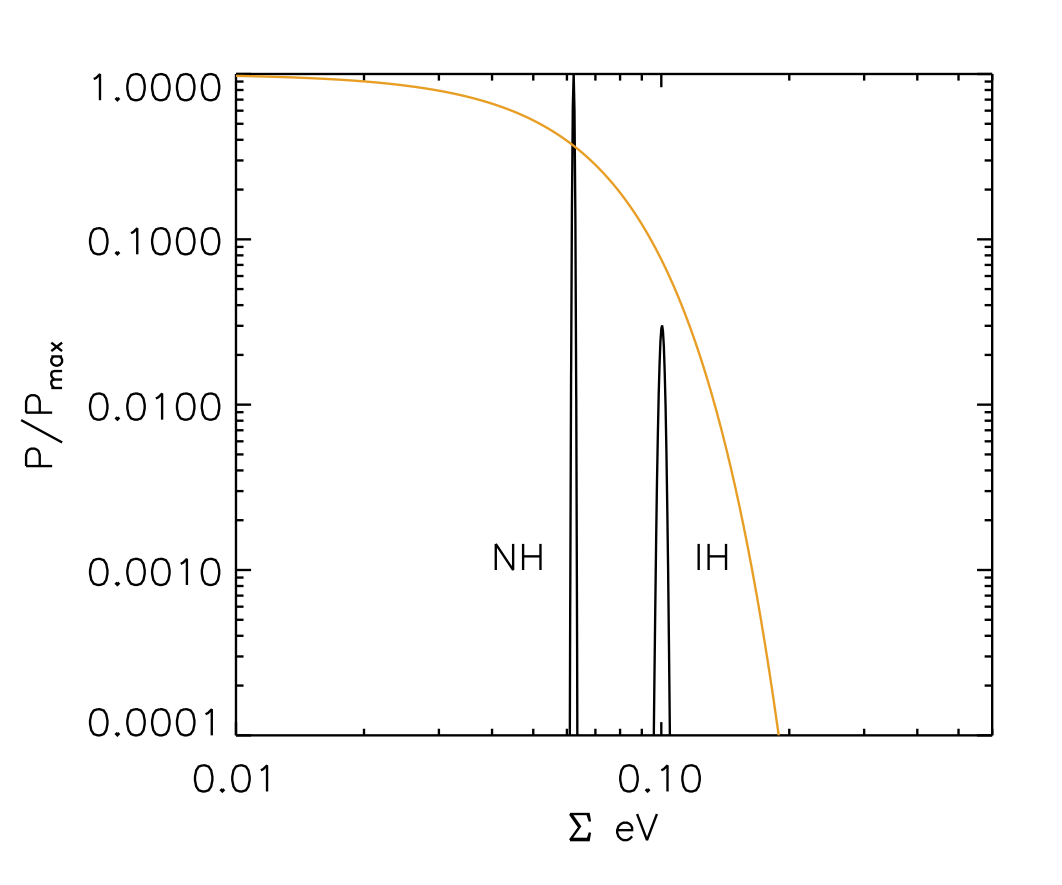}
\caption{Visualization of current constraints. On the top two panels, the two thin (blue, NH and red, IH) ridges in the 3D space of $m_1,m_2,m_3$   represent the mass splitting constraints (95\% CL) from oscillation experiments in eq.~\eqref{eq:splittings}. For the NH ($m_1,m_2,m_3$) corresponds to $(m_{\rm L},m_{\rm M},m_{\rm H})$, while for the IH it corresponds to $(m_{\rm M},m_{\rm H},m_{\rm L})$. For visualization purposes the NH ridge is shown only for $m_{\rm L}>10^{-3}$ eV. In the uppermost panel only oscillation constraints are used, while in the middle panel a cosmology constraints of $\Sigma<0.1$ eV is imposed; note the significant decrease of the allowed IH region and the different z-axis range. The bottom panel shows the normal and inverted  hierarchy oscillations likelihood probability distributions ($P/P_{\rm max}$) for the sum of the masses, keeping the mass of  the lightest neutrino fixed at $m_{\rm L}=3\times 10^{-3}$ eV and cosmology constraint $\Sigma_{\rm cosmo}<0.089$ eV (95\% C.L., orange).}
\label{fig:oldnewfig1}
\end{figure}

\section{Results}
\label{sec:results}
We begin by presenting a visualization of the parameter space available for the two hierarchies,  given the updated constraints  illustrated in figure~\ref{fig:oscillhierarchy} where the probability distribution for the sum of the masses $\Sigma$ is shown on the right panel. The black lines correspond to the oscillation constraints for the normal and inverted hierarchies, as indicated in the labels, and the grey bands represent the $5\sigma$ uncertainty regions. The vertical lines show the various $\Sigma_{\rm cosmo}$ constraints.

However, the (e.g., 95\% C.L.) regions allowed by the oscillation experiments are actually in the 3D space of $(m_1,m_2,m_3)$; recall that for the NH this corresponds to $(m_{\rm L},m_{\rm M},m_{\rm H})$ and for the IH this corresponds to $(m_{\rm M},m_{\rm H},m_{\rm L})$. Therefore, in figure~\ref{fig:oldnewfig1} we show the 3D locus of the hierarchies as allowed by current oscillation data (top panel) and after imposing a cosmological constraint of $\Sigma < 0.1$ eV (middle panel); note the significant decrease of the allowed region for the IH. The bottom panel shows the probability distribution of the sum of the masses given the oscillations measurements for both the NH and IH and fixing the lightest neutrino to $m_{\rm L}=3\times 10^{-3}$ eV, and the effect of the cosmological constraint (orange line) at further decreasing the allowed region for the IH.

\begin{figure}[tbp]
\centering
\includegraphics{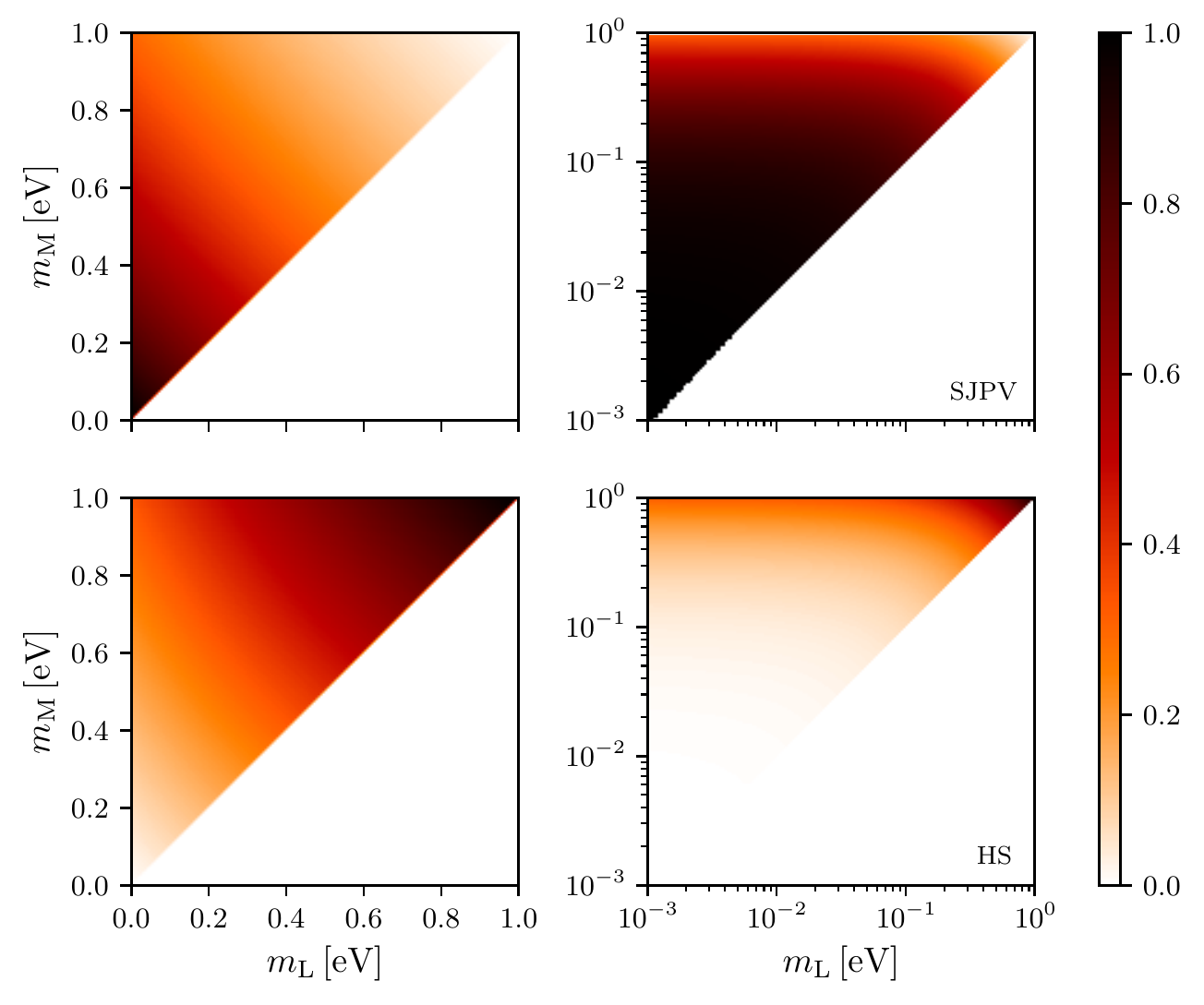}
\caption{Comparison of the SJPV (logarithmic hierarchical) and HS (objective Bayesian) prior density $P({\bf m})$: top panels show the SJPV prior probability before imposing oscillations constraints for an illustrative choice of hyper-parameters  $\log \mu=-1.5$, $\sigma=1$ in the $m_{\rm L}$-$m_{\rm M}$ plane, for a fixed $m_{\rm H}$, while bottom panels show the corresponding probability density for the HS prior. Shown for axes in linear (left) or logarithmic (right) scale. The color scale is linear with black at the maximum ($P/P_{\rm max}=1$) and white at the minimum. The lower triangle is excluded by the condition $m_{\rm L}\le m_{\rm M}\le m_{\rm H}$. The HS prior decreases for low $m_{\rm M}$ and $m_{\rm L}$ and, for a fixed value of $m_{\rm L}$, is minimal when $m_{\rm M}=m_{\rm L}$; for the SJPV prior, the opposite is true.}
\label{fig:priors}
\end{figure}
\begin{figure}[tbp]
\centering
\includegraphics{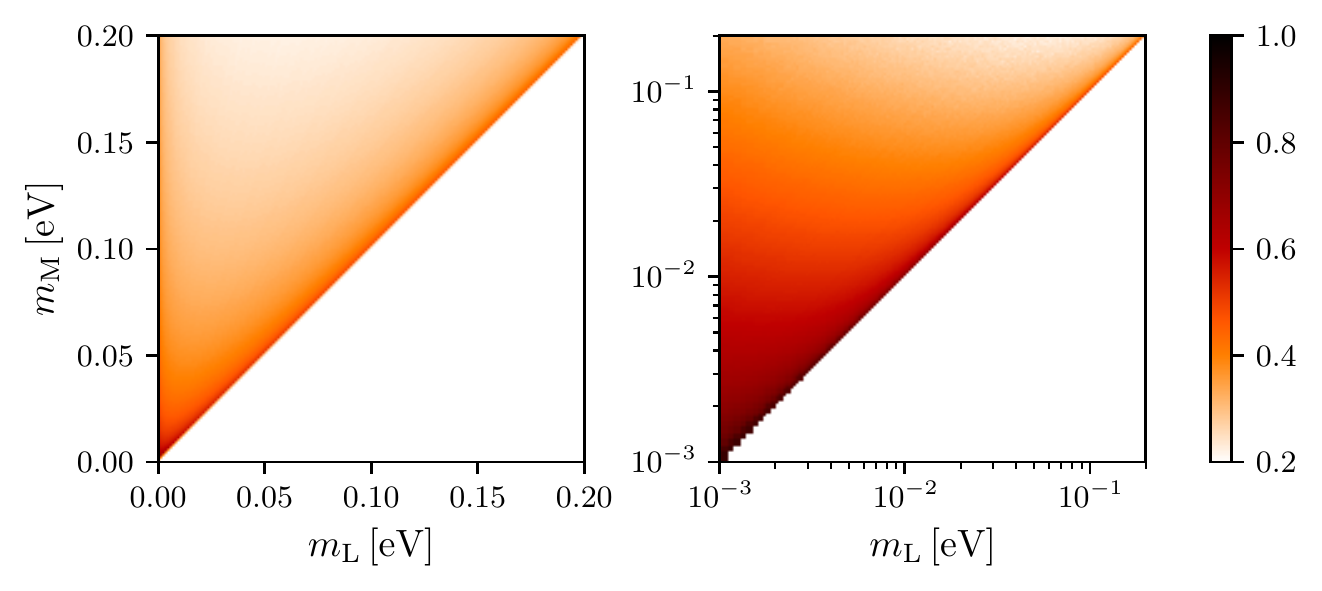}
\caption{SJPV hyperprior density $P({\bf m})$ once constraints from oscillation data are included, marginalized over the hyper-parameters $\mu$ and $\sigma$ and over $m_{\rm H}$; no constraints on $\Sigma$ are imposed. Axes in linear (left) and logarithmic (right) scale, with linear color-scale artificially cut at 0.2 to increase contrast. 
Here it is evident that the SJPV hierarchical prior, which encodes information from oscillations, differs drastically from HS, favoring low masses.
}
\label{fig:priors2SJVP}
\end{figure}

It is very illustrative to visualize the priors of the two different approaches and compare them. In figure~\ref{fig:priors}, the top panels show the logarithmic hierarchical (SJPV) prior probability density, $P({\bf m})$, in the $m_{\rm L}$-$m_{\rm M}$ plane  for a fixed $m_{\rm H}=1\, \rm{eV}$ and before considering the oscillations data. We show this for a representative choice of hyper-parameters $\log \mu=-1.5$, $\sigma=1.0$, noting that while this is a somewhat low value of $\sigma$ (see discussion in appendix~\ref{sec:appendixBHM}), a higher value would be less illustrative as the distribution on the right panel would look very uniform. The specific choice of value for $m_{\rm H}$ does not affect the distribution, provided that $m_{\rm H} \ge m_{\rm M}$. The prior distributions are shown with axes in both linear (left) and logarithmic (right) scales. The bottom row panels show the corresponding probability density for the objective Bayesian (HS) prior. In this case, changing the value of $m_{\rm H}$ simply rescales the values on the $m_{\rm L}$ and $m_{\rm M}$ axes. The HS prior decreases for smaller values of $m_{\rm M}$ and $m_{\rm L}$ and, for a fixed value of $m_{\rm L}$, is minimal when $m_{\rm M}=m_{\rm L}$. On the other hand, the SJPV prior (before any oscillation data) does exactly the opposite; it increases for low $m_{\rm M}$ and $m_{\rm L}$ and, for a fixed value of $m_{\rm L}$, is maximal when $m_{\rm M}=m_{\rm L}$.

In figure~\ref{fig:priors2SJVP} we demonstrate how the SJPV hyperprior changes once the hyper-parameters are constrained by the inclusion of the oscillations data. It is evident that once oscillation constraints are taken into account, the resulting hierarchical prior becomes drastically different. The oscillations-informed hyperprior heavily favors either $m_{\rm L}\simeq m_{\rm M}$ (NH, diagonal ridge) or $m_{\rm L}\ll m_{\rm M}$ (IH). Even along these directions this prior favors low masses and the smooth cutoff at high masses becomes more stringent with subsequent improvements on the $\Sigma$ limits. 

\begin{figure}[tbp]
\centering
\includegraphics{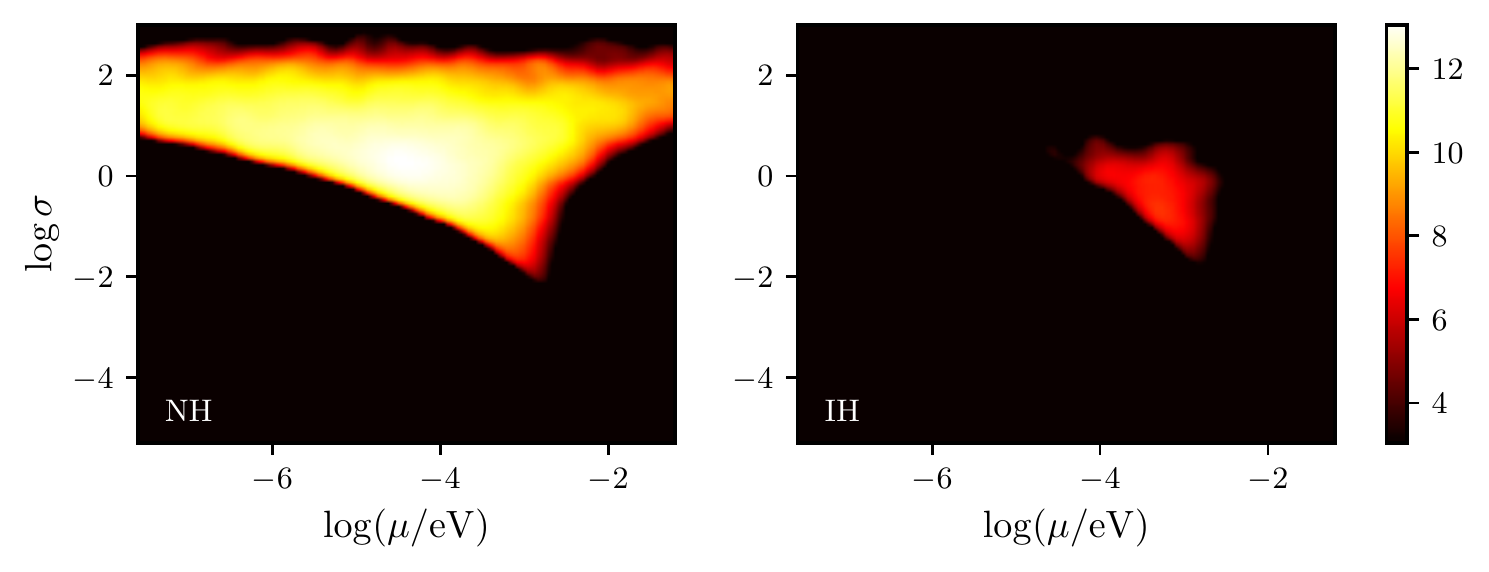}
\caption{Normal (left) and Inverted (right) hierarchy marginal log-likelihood distributions for the hyper-parameters of the hierarchical prior for a cosmology mass limit of $\Sigma_{\rm cosmo}<0.089$ eV (95\% C.L.). The colorbar convention used here is that of~\cite{Fergus}, for easier comparison.}
\label{fig:cubed_graph}
\end{figure}

The probability density of the hyper-parameters $\mu$ and $\sigma$ for the logarithmic hierarchical SJPV prior is shown in figure~\ref{fig:cubed_graph} for the normal (left) and inverted (right) hierarchies after imposing cosmological constraints. A direct comparison with figures 2 and 3 of ref.~\cite{Fergus} shows how the hyper-parameter values change as the bound on $\Sigma$ becomes more stringent, progressively disfavoring the hyper-parameter space for the IH more than for the NH.

\begin{table}[tbp]
\begin{center}
\begin{tabular}{@{}llllll@{}}
\toprule
 & \multicolumn{2}{c}{SJPV}    &  & \multicolumn{2}{c}{HS}  \\ \cmidrule(lr){2-3} \cmidrule(l){5-6} 
 & \multicolumn{1}{c}{$\kappa_{{\rm no}\Delta_{\chi^2}}$} & $K_{\rm SJPV}$ (using $\Delta{_\chi^2}$) &  & \multicolumn{1}{c}{$\kappa_{{\rm no}\Delta_{\chi^2}}$} & $K_{\rm HS}$ (using $\Delta^2_\chi$) \\ \midrule
 $\Sigma <2.85$ eV (95\%) [A]  & 3.8  & 124 &  & 1.0 & 33.1 \\
$\Sigma < 0.8$ eV (95\%) [B]  & 5.3  & 175 &  & 1.2 & 38.1 \\
$\Sigma < 0.12$ eV (95\%) [C] & 49   & 1607 &  & 3.6  & 118  \\
$\Sigma < 0.102$ eV (95\%) [D] & 87  & 2861 &  & 4.3  & 143 \\
$\Sigma < 0.099$ eV (95\%) [E] &  88 & 2890  &  & 4.7 &  156 \\
$\Sigma < 0.089$ eV (95\%) [F]&  138 & 4553  &  & 7.4 & 244 \\ \bottomrule
\end{tabular}
\end{center}
\caption{Bayesian evidence ratios for the NH vs IH with SJPV and HS priors. The $\Sigma$ constraints  assume zero-centred Gaussian distributions and 95\% C.L. limits as follows: A=KATRIN current limit~\cite{katrin21},
B=KATRIN sensitivity limit~\cite{katrin21}, C=Planck+BAO~\cite{eBOSSresults}, D=Planck+BAO+RSD~\cite{eBOSSresults}, E=Planck+BAO+RSD+SNe~\cite{eBOSSresults} F=Planck+BAO+Ly-$\alpha$~\cite{1911.09073}. The sensitivity to the interpretation of the $\Sigma_{\rm cosmo}$ constraints is explored in table~\ref{tab:evidenceconvention}.}
\label{tab:evidence}
\end{table}

We are now ready to interpret the Bayesian evidence results presented in table~\ref{tab:evidence} for both the SJPV and SH priors as a function of the $\Sigma$ upper limit, assuming a zero-centered Gaussian distribution truncated at $\Sigma =0$. The error in the calculation of the log-evidence values is estimated to be around $0.05$--$0.1$, hence negligible for our purposes. Results with or without the Super-Kamiokande atmospheric (SK-atm) data are reported in section~\ref{sec:sensitivity} for completeness. However, we see no reason to exclude this data set, hence our baseline results discussed here always include it. 

The new oscillation constraints do not significantly alter the evidence compared to the 2018 situation. In the HS case, compared with ref.~\cite{HeavensEvidence}, the major effect is due to the increased $\Delta\chi^2$ between the two hierarchies which has risen from  $\Delta\chi^2 = 0.83$ in 2018 to  $\Delta\chi^2 =7.0$ now, increasing the contribution to the HS evidence ratio from the $\Delta\chi^2$ alone (which is an overall multiplicative normalization factor) from $\kappa_{\Delta\chi^2}=1.5$ to $\kappa_{\Delta\chi^2}=33$. Ref.~\cite{Fergus}, using the SJPV prior, reported evidence ratios neglecting the contribution from the $\Delta \chi^2$ as it had a negligible effect for the oscillations measurements adopted in that work.  One can also appreciate that the increasingly stringent mass limits have only a mild effect on the evidence ratio for the HS prior choice, but a much more significant effect for the SJPV prior.
 
By construction both priors, in the absence of oscillations data, do not favor one hierarchy over another\footnote{See discussions in refs.\cite{Fergus, HeavensEvidence}. Briefly, the priors do not distinguish the three mass eigenstates, --this is called exchangeability, see appendix \ref{sec:appendixBHM}-- hence do not favor one hierarchy over another before the data arrives.}. Once the oscillations data are taken into account, but before considering any constraint on $\Sigma$, both priors give very similar posterior odds, which now are driven by the $\Delta \chi^2$. Recall that the oscillation experiment measurements of the mass-squared splittings in eq.~\eqref{eq:splittings} have not changed dramatically since 2018, although the value of the $\Delta \chi^2$ has. As long as the constraints on the sum of the masses are weak -- e.g. $\Sigma_{\beta} <6.9$ eV (adopted in SJPV) or $\Sigma_{\rm cosmo} <1.5$ eV (adopted in HS) -- as in 2018, the evidence ratio $\kappa_{{\rm no} \Delta \chi^2}$ is inconclusive ($<3$) for both prior choices, although the $\kappa_{\Delta \chi^2}$  contribution now boosts the overall posterior odds  for the NH by a factor 33. However, the situation changes drastically as the bounds on $\Sigma$ tighten: in the SJPV case, the evidence is much more sensitive to the $\Sigma$ constraints and the evidence for the NH increases much more dramatically once the total mass limit crosses $\Sigma<0.1$ eV. Regardless of the prior choice, the odds ratio for NH is always $\gtrsim$ 100 when including the latest cosmological constraints $\Sigma_{\rm cosmo}$ which disfavour values of $\Sigma$ above 0.1 eV.

\begin{table}[tbp]
\begin{center}
\begin{tabular}{|c|c|}
\hline
K& Strength of evidence\\
\hline
1--3.2 & Not worth more than a bare mention\\
3.2--10& Substantial \\
10--100 &Strong \\
$>$100 & Decisive\\
\hline
\end{tabular}
\end{center}
\label{tab:KR95}
\caption{Kass and Rafetry 1995~\cite{KR95}  qualitative interpretation of Bayesian evidence ratios. This scale is empirically calibrated.   $K$ is a Posterior Bayesian odds ratio, the relative plausibility of two models in light of the data,  and should not be interpreted as number of sigmas or a p-value.}
\end{table}

The Bayesian evidence ratio values or odds can be interpreted qualitatively according to e.g. the Kass and Raftery~\cite{KR95} scale reported in table~\ref{tab:KR95}, which indicates that in all cases for both choices of priors the evidence ratio is always ``Strong" or ``Decisive"  when including the $\kappa_{\Delta_\chi^2}$ normalization.
Independently of the prior choice the evidence for the NH is now ``Decisive''. For the HS prior, compared to the results from 2018 when the evidence was weak, this is driven by the $\Delta \chi^2$ obtained from the global fit of oscillations data. For the SJPV prior  the evidence moves from ``Strong" to ``Decisive" even without accounting for the $\Delta \chi^2$ contribution, and is driven by the improved cosmological limit on the sum of the masses.

\begin{figure}[tbp]
\centering
\includegraphics{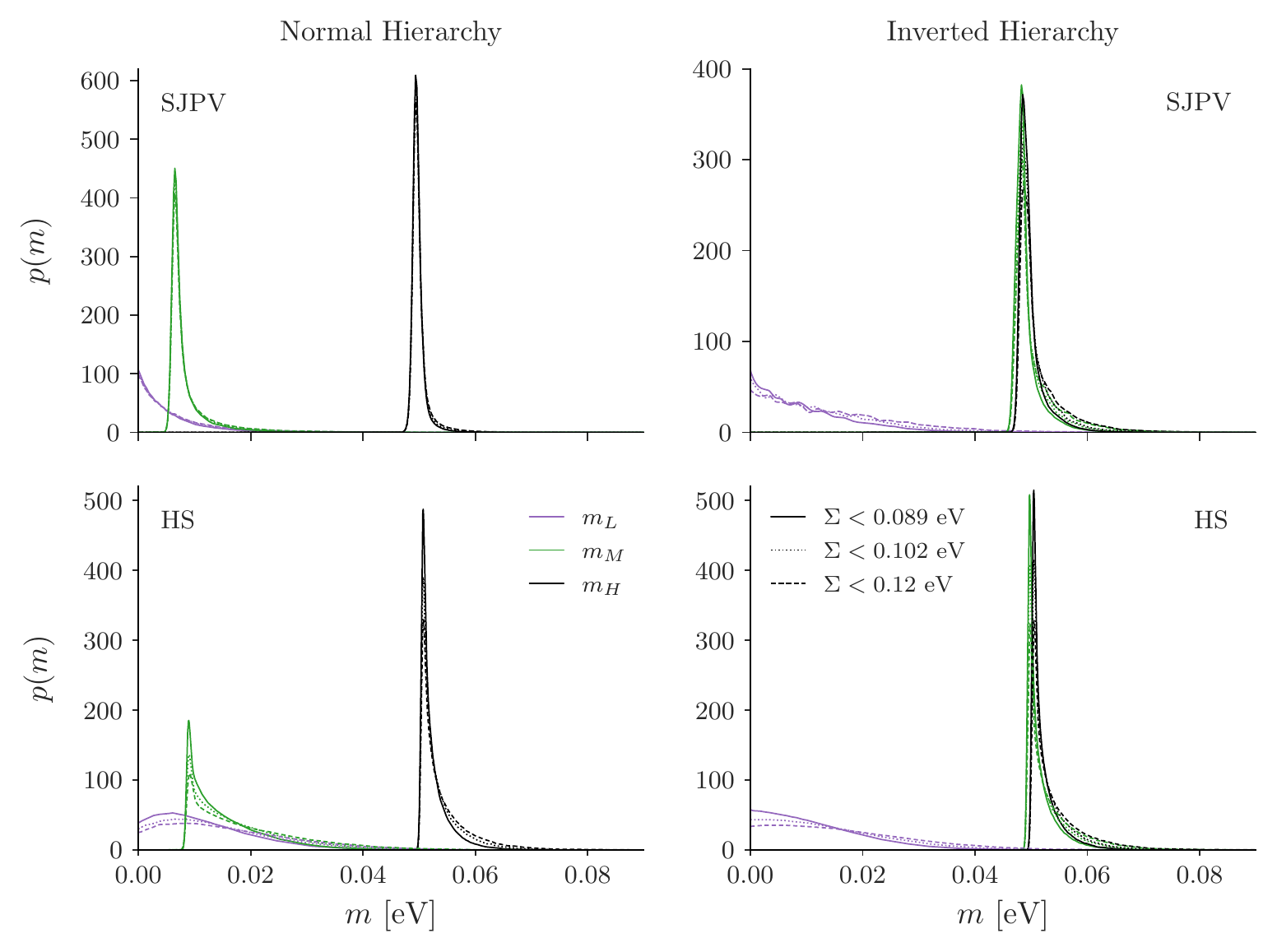}
\caption{Posterior distributions $p(m_i)=P(m_i|D,M)$ for the individual masses and sensitivity to the adopted $\Sigma$ limit for the NH (left) and the IH (right). Upper panels show the result with the SJPV prior, bottom panels the HS prior.}
\label{fig:x cubed graph1}
\end{figure}

\begin{figure}[tbp]
\centering
\includegraphics{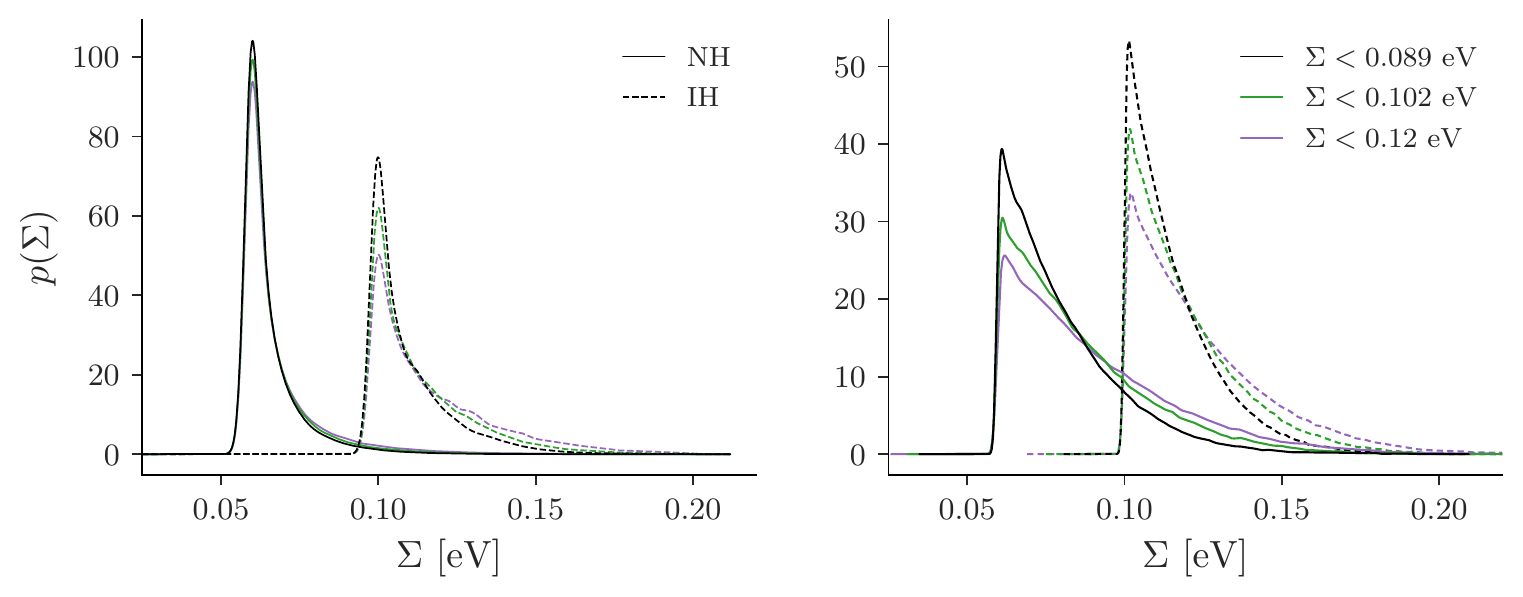}
\caption{Comparison of the posterior distributions for the sum of the masses with the SJPV prior (left) and the HS prior (right). Dashed lines denote the IH, solid lines the NH. The weighted combined distribution for the two hierarchies is indistinguishable from that of the NH, because of the overwhelming odds.}
\label{fig:x cubed graph2}
\end{figure}

Finally, we report the posterior distributions for the individual masses and the sum of the masses for both NH and IH according to the two prior choices in figures~\ref{fig:x cubed graph1} and \ref{fig:x cubed graph2}. The distributions of the individual masses are very similar for the two prior choices: once a hierarchy is chosen and the degenerate regime (where all the masses are much bigger than the splittings) is excluded by the constraints on $\Sigma$, the individual masses are determined by the oscillations constraints on the mass splittings (eq.~\eqref{eq:splittings}). The probability distribution for the sum of the masses of course differs under NH or IH assumptions, 
but when taking the weighted sum of the two components the combined probability distribution for $\Sigma$ is virtually indistinguishable from that for the NH for either prior choice. This reflects the fact that the evidence prefers the NH over the IN with odds greater than 100:1. 
The effect of the stronger $\Sigma$ limit can be appreciated by comparing this figure with figure 8 of ref.~\cite{Fergus}, where a limit of $\Sigma<6.9$ eV was adopted.

\section{Sensitivity analysis}
\label{sec:sensitivity}

The above results are quite robust to variations over the baseline setup of section~\ref{sec:results}; in a few cases these variations affect the results quantitatively but not qualitatively, which we discuss below.
In ref.~\cite{gonzalez-garciaglobal21} results are also reported without the SK-atm constraints, in which case the constraints on $\Delta m_{3\ell}^2$ in eq.~\eqref{eq:splittings} become
\begin{equation}
 \label{eq:m13_noSKatm}
 \Delta m_{3\ell}^2 = m_3^2 - m_\ell^2 = 
\begin{cases}
  \phantom{-}   2.515  \, (\pm 0.028) \times 10^{-3} \mathrm{eV}^2   ,& \text{(NH) }  \\
     -2.498 \, (\pm 0.029) \times 10^{-3} \mathrm{eV}^2        ,& \text{(IH)}
\end{cases}
\,\,\, (68.4\% {\rm CL})\,.
\end{equation}

The evidence ratios both with and without inclusion of SK-atm data for both prior choices are reported in table~\ref{tab:SJPV_HS_evidence}. For each case, we report the resulting odds before and after including the $\Delta \chi^2$ normalisation from oscillations data. While the SK-atm data does not affect much the $\kappa_{{\rm no}\Delta \chi^2}$, the effect on the $\kappa_{\Delta \chi^2}$ is major, ranging from 3.7 (no SK-atm) to 33 (with SK-atm). Both cases are reported for completeness, but we see no reason to exclude the SK-atm data set or the $\Delta \chi^2$ normalisation, and hence our baseline results include it.

\begin{table}[t]
\centering
\resizebox{\textwidth}{!}{%
\begin{tabular}{@{}lcclcc@{}}
\toprule
                               & \multicolumn{2}{c}{NuFITv5.1 (no SK-atm data)}                                                                                     &  & \multicolumn{2}{c}{NuFITv5.1 (with SK-atm data)}                                                                                 \\ \midrule
SJPV                           & $\kappa_{\rm{no}\Delta_{\chi^2}}$                     & \multicolumn{1}{l}{$K_{\rm SJPV}\, ({\rm with} \, \Delta{_\chi^2} = 2.6$)} &  & $\kappa_{{\rm no}\Delta_{\chi^2}}$                     & \multicolumn{1}{l}{$K_{\rm SJPV}\, ({\rm with}\, \Delta^2_\chi = 7.0$)} \\ \cmidrule(l){2-6} 
$\Sigma < 0.8$ eV (95\%) [A]   & 4.9                                                   & 18.1                                                                       &  & 5.3                                                    & 174.8                                                                   \\
$\Sigma < 0.12$ eV (95\%) [B]  & 50.2                                                  & 219.4                                                                      &  & 48.7                                                   & 1607.4                                                                  \\
$\Sigma < 0.102$ eV (95\%) [C] & 87.3                                                  & 322.9                                                                      &  & 86.7                                                   & 2861.5                                                                  \\
$\Sigma < 0.099$ eV (95\%) [D] & 91.2                                                  & 337.6                                                                      &  & 87.5                                                   & 2888.6                                                                  \\
$\Sigma < 0.089$ eV (95\%) [E] & 131.2                                                 & 485.3                                                                      &  & 138.0                                                  & 4553.8                                                                  \\ \midrule
HS                             & \multicolumn{1}{l}{$\kappa_{\rm{no}\Delta_{\chi^2}}$} & \multicolumn{1}{l}{$K_{\rm HS}\, ({\rm with} \, \Delta{_\chi^2} = 2.6$)}   &  & \multicolumn{1}{l}{$\kappa_{{\rm no}\Delta_{\chi^2}}$} & \multicolumn{1}{l}{$K_{\rm HS}\, ({\rm with}\, \Delta^2_\chi = 7.0$)}   \\ \cmidrule(l){2-6} 
$\Sigma < 0.8$ eV (95\%) [A]   & 1.1                                                   & 4.0                                                                        &  & 1.2                                                    & 38.1                                                                    \\
$\Sigma < 0.12$ eV (95\%) [B]  & 3.4                                                   & 12.4                                                                       &  & 3.6                                                    & 118.4                                                                   \\
$\Sigma < 0.102$ eV (95\%) [C] & 4.1                                                   & 15.2                                                                       &  & 4.3                                                    & 143.4                                                                   \\
$\Sigma < 0.099$ eV (95\%) [D] & 4.6                                                   & 16.8                                                                       &  & 4.7                                                    & 156.1                                                                   \\
$\Sigma < 0.089$ eV (95\%) [E] & 6.4                                                   & 23.3                                                                       &  & 7.4                                                    & 244.2                                                                   \\ \bottomrule
\end{tabular}%
}
\caption{Bayesian evidence ratios with the SJPV and HS priors. As ref.~\cite{gonzalez-garciaglobal21} present results with and without including Super-Kamiokande atmospheric constraints, here we also report both cases for completeness. The upper limits on $\Sigma$ are follows: A=KATRIN sensitivity limit
~\cite{katrin21}, B= Planck+BAO\cite{eBOSSresults}, C=Planck+BAO+RSD~\cite{eBOSSresults}, D=Planck+BAO+RSD+SNe~\cite{eBOSSresults} E=Planck+BAO+Ly-$\alpha$~\cite{1911.09073}.}
\label{tab:SJPV_HS_evidence}
\end{table}

The sensitivity of the posterior odds to the choice of interpretation of the cosmological constraints on $\Sigma$ is presented in table~\ref{tab:evidenceconvention}. Clearly, for a fixed 95\% C.L. limit, there is some sensitivity to the assumed shape (i.e., location of the peak) of the probability distribution, but the evidence is always ``Strong" or ``Decisive" once the $\Delta \chi^2$ is included. Excluding SK-atm on the other hand lowers the $\kappa_{\Delta\chi^2}$ by almost a factor of 10, from $33$ to $3.7$, but still larger than the 2018 (HS adopted) value of $\kappa_{\Delta\chi^2}=1.5$.

\begin{table}[tbp]
\centering
\begin{tabular}{@{}lcc@{}}
\toprule
\multicolumn{1}{c}{Dataset}                          & $\kappa_{\rm SJPV}$ (K$_{\rm SJPV}$, SJPV prior) & $K_{\rm HS}$ (HS prior) \\ \midrule
Splitting constraints ($\Sigma<6.9$ eV)               & 2.6\, (85.8)                             & 1.0\, (33)          \\
$\Sigma<0.8$ zero-centred Gaussian                    & 5.3 \, (175)                             & 38                  \\
$\Sigma_{\nu} < 0.102 $ Gaussian centred at -0.026 eV & 74 \, (2442)                             & 138                 \\
$\Sigma_{\nu} < 0.102$ zero-centred Gaussian          & 87 \,  (2860)                            & 143                 \\
$\Sigma_{\nu} < 0.099$ zero-centred Gaussian          & 88\, (2900)                              & 156                 \\
$\Sigma_{\nu} < 0.099$  Gaussian centered at 0.04 eV  & 43\, (1419)                              & 92                  \\
$\Sigma_{\nu} < 0.089$ zero-centred Gaussian          & 138 \, (4550)                            & 244                 \\ \bottomrule
\end{tabular}
\caption{
    Bayesian evidence ratios for the baseline case but using different interpretations of the $\Sigma_{\rm cosmo}$ constraints (in eV, 95\% C.L.). The rationale for centering the distribution at $-0.026$ eV is that this is suggested by ref.~\cite{eBOSSresults}; the other plausible alternative to centering the distribution at zero is to center the distribution at the minimum  value allowed by oscillations $\Sigma \sim 0.04$ eV.} 
    \label{tab:evidenceconvention}
\end{table}

\section{Conclusions}

Adopting the latest constraints on the sum of neutrino masses from cosmological observations and on the mass-squared splittings from ground-based oscillation experiments, we computed the Bayesian evidence of a given neutrino hierarchy. Our main conclusion is that new data prefer ``Strongly" if not ``Decisively" the normal hierarchy, with odds over 100:1 ( from 140:1 to over 2000:1 depending on the prior choice).  Compared to similar analyses presented with older oscillation experiments constraints, this result is driven by two main effects: the increased $\Delta\chi^2$ from global fits to  oscillations data  and  the more stringent limits on the sum of neutrino masses from cosmology (obtained under the assumption of a $\Lambda$CDM model).
It is well known that the evidence calculation is sensitive to the prior choice. To demonstrate the robustness of this result we have considered two widely different priors: a set of hyperpriors in logarithmic space on the individual neutrino masses and the so-called ``Objective Bayesian'' reference prior, linear in the individual masses, considered in ref.~\cite{objectivebayesian} to make the prior as minimally informative as possible to the oscillations measurements. These two cases are at the extreme of possibilities for prior choices, and therefore bracket the range of results achievable with other (physical) choices of priors. While the exact numerical value for the Evidence depends on the specific choice of prior, and the specific limit on the sum of the masses,  for the latest cosmological constraints\footnote{In a otherwise standard $\Lambda$CDM  cosmological model. } on the sum of the masses, the evidence remains   ``Decisive" across very different prior choices. For the HS prior, cosmological constraints on the sum of the masses are crucial to push the Evidence above the ``decisive" threshold. It is useful to bear in mind that  cosmological constraints are well known to be model dependent: constraints on $\Sigma_\nu$ degrade for non-standard cosmological models.
 There are examples in the literature (e.g.,~ \cite{GariazzoMena19, Hergt, Mahony} and refs. therein) where  the choice of the prior leads to an evidence for the normal hierarchy  weaker than reported here. This is achieved  at the cost of having different prior distributions for the different individual masses. Physically this would imply that the three masses do not share a common origin and therefore that  there are three different  (yet undiscovered) mechanisms of mass generation for the different mass eigenstates. 

We have also predicted the probability distribution for values of the individual neutrino masses and of the sum of the masses that future experiments should measure.

Our findings update and corroborate the previous results of ref.~\cite{Fergus} and significantly increase the Bayesian evidence for the normal hierarchy. This has important consequences for neutrino physics and experiments that search for neutrino-less double beta decay. Current efforts aim at the ton-scale active material experiments, which are sensitive to the 10 meV scale in the effective neutrino mass of this decay. In light of the results of this work, the motivation for covering the inverted hierarchy scenario is lost and the focus should shift to maximising the fraction of the normal hierarchy parameter space covered by a given experiment. Moreover, our results increase the incentive for neutrino-less double beta decay experiments that would further cover the parameter space for the NH, either due to their ability to scale up the mass of the decaying nucleus or because the choice of nucleus is favoured by a smaller decay half-life.

\acknowledgments

We thank Alan Heavens for stimulating discussions, healthy and exquisitely professional scientific and interpretive disagreement, and for an open dialogue even with dissenting opinions, which should be more common than it actually is in the scientific discourse. We recognize that the exchange helped shape this paper in its present form. 
KS acknowledges support from the European Union’s Horizon 2020 research and innovation programme under the Marie Sk\l{}odowska-Curie grant agreement No.~713673. KS is also supported by an INPhINIT fellowship from ``la Caixa" Foundation (ID 100010434), grant code LCF/BQ/DI17/11620047.
Funding for this work was partially provided by project PGC2018-098866-B-I00 \\ MCIN/AEI/10.13039/501100011033 y FEDER “Una manera de hacer Europa”, and the “Center of Excellence Maria de Maeztu 2020-2023” award to the ICCUB (CEX2019-000918-M funded by MCIN/AEI/10.13039/501100011033).
LV acknowledges support of  European Union's Horizon 2020 research and innovation programme ERC (BePreSysE, grant agreement 725327).

\appendix

\section{Bayesian hierarchical modelling}
\label{sec:appendixBHM}
A key difference between the analysis of ref.~\cite{Fergus} and other works is that the prior is hierarchical in nature. Hierarchical priors were first explored in the 1960s, but their impact was not fully realised until the 21st century. This is mostly due to the increase in computing power, as computing the evidence from these priors is computationally expensive.
 Rather adopting a single pre-determined prior, of fixed position and breadth, one adopts instead a hyperprior, that is one  introduces   an entire family of priors and effectively marginalises over them. Without repeating the arguments presented in ref.~\cite{Fergus}, the rational can be summarized  as follows:   Hyperpriors are thus  more flexible and as a result of marginalizing over the hyper-parameters describing this family of priors,  the posterior is  less sensitive to the prior itself. 
To appreciate how hierarchical priors can help improve measurements, consider the following scenario. Alice and Bob are the first astronauts ever to set foot on an exoplanet. Upon exiting their spacecraft, they soon come across a new life form, a species they call snargs. In order to collect data on the nature of snargs, Alice and Bob perform various measurements on the individual creatures. One by one, Alice and Bob begin measuring how much each adult snarg weighs. The first specimen weighs $9.8$~kg, while the second is found to weigh $9.7$~kg. For these measurements, the accuracy of the weighing device is one part in a thousand, and the uncertainties in the separate measurements can be assumed to be statistically independent (i.e., not due to a calibration error in the weighing device).

After some time, Alice and Bob have successfully measured the masses of 99 different snargs. The mean mass of the snargs they have sampled so far is $9.6$~kg, with a standard deviation of $0.2$~kg. However, they are not so fortunate with the 100th snarg. Activity from a nearby star causes an electrical storm to engulf the planet, resulting in both of their weighing devices to malfunction when measuring the 100th snarg. The weighing device returns a measurement of $25$~kg, along with an uncertainty of $20$~kg. What is the true mass of the 100th snarg? Alice and Bob are both Bayesians, but they adopt different priors on their snarg masses. Alice adopts independent uniform priors on each of the snargs. Her posterior belief on the mass of the 100th snarg is therefore $25$~kg with a standard error of $20$~kg. Bob adopts a different approach. Bob  assumes that the mass of every snarg (including the 100th they weigh) comes from the same underlying distribution, and adopts a hyerarchical prior  which is described by  two hyperparameters: the mean and the standard deviation. Bob marginalizes over the hyperparameters values, in the usual way. The precise measurements on the first 99 snargs effectively constrain the values of the hyperparameters.  By using a hierarchical model on the mass of the snargs, Bob's posterior belief is that the mass of the 100th snarg is $9.6$~kg with a standard error of $0.2$~kg. Alice's and Bob's beliefs in the mass of the 100th snarg differ in their precision by a factor of one hundred. 

Hyperpriors are more flexible than standard priors and make the posterior less sensitive to the prior itself. Following ref.~\cite{Fergus}, we work in a five-dimensional parameter space, consisting of the three neutrino masses indicated by the vector ${\bf m}={m_1, m_2, m_3}$, and two hyperparameters $\mu$ and $\sigma$. This takes the form
\begin{equation}
    P(D, {\bf m}, \mu, \sigma |M) = P(D|{\bf m}, M)P({\bf m}|\mu, \sigma, M) \pi (\mu, \sigma | M)
\end{equation}
where $D$ represents the data vector and $M$ the Model or hypothesis i.e., normal or inverted hierarchy, and $\pi (\mu, \sigma |M) $ is the hyperprior. 

The prior must reflect our state of belief \emph{before} the data arrived. As the three masses are indistinguishable before the (oscillation) data arrives,  our prior must (and does by construction, as the three masses are drawn from a common prior) reflect this symmetry. This feature is also known as exchangeability, which naturally arises if the three masses share a common origin, a hypothesis that is adopted in most particle physics models for the neutrino masses. 
The relation between exchangeability and hierarchical prior is further explored in  \cite{Bernardo, Finetti31, BernardoSmith09}, but we will not dwell on this here.
Following ref.~\cite{Fergus} we adopt a hierarchical lognormal prior $P(\log {\bf m}|\mu, \sigma, M)\sim {\cal N}(\log \mu,\sigma)$, where ${\cal N}$ denotes the Normal (Gaussian) distribution with mean and standard deviation controlled by the two hyperparameters $\mu$ and $\sigma$. Here $\mu$ represents the median mass value associated with the lognormal distribution, while $\sigma$ denotes the standard deviation of the Gaussian in log space. The hyperprior imposed on our hyperparameters is given by 
$\pi (\log \mu, \log \sigma | M) \sim {\cal U}(\log \mu_{\rm min}, \log \mu_{\rm max}) {\cal U}(\log \sigma_{\rm min}, \log \sigma_{\rm max})$, where ${\cal U}$ denotes a uniform distribution between the maximum and minimum bounds of $\mu_{\rm min}=5 \times 10^{-4}$ eV, $\mu_{\rm max}=0.3$ eV and $\sigma_{\rm min}=5\times 10^{-3}$ eV, $\sigma_{\rm max}=20$ eV.

\section{Objective Bayesian prior}
\label{sec:appendixOBP}
For completeness, we report here the details on our implementation of the HS prior of ref.~\cite{objectivebayesian}.
The approach in ref.~\cite{objectivebayesian} is to construct a prior that contributes minimal information to the analysis;
or, in other words, the prior that maximises the expected information gain from the data. 
A `reference prior' (see e.g.~\cite{reference_prior}) is used, which maximises the missing information between the posterior and prior as data arrives, in a mathematically well-defined sense. In the case of the neutrino hierarchy problem, where the likelihood has achieved asymptotic normality, the reference prior corresponds to the Jeffreys prior and so uses the Fisher information as information measure. 

The natural parameters to use then are the small and large squared mass splittings, $\phi$ and $\psi$ respectively,\footnote{Ref.~\cite{HeavensEvidence} uses  an alternative  definition  for the large mass splitting, but this is unimportant for this application.} and the sum of the masses $\Sigma$.
With the precision of current measurements, all three parameters ($\phi, \, \psi, \, \Sigma$) can be treated as linear and the standard deviation as constant, which results in uniform Jeffreys priors for each of the parameters ($\phi, \, \psi, \, \Sigma$). Due to the independence of the datasets, the total reference prior is the product of the three reference priors for $\phi, \, \psi, \, \Sigma$. Hence these are taken to be uniform, excluding non-physical values. This prior is improper so a maximum value for $\Sigma$ must be taken, which HS take to be 1.5 eV.
This prior can then be transformed to the reference prior in the ($m_{\rm L}, \, m_{\rm M}, \, m_{\rm H}$) parameterisation using the Jacobian, such that

\begin{align}
\begin{split}
 J(m_{\rm L}, m_{\rm M}, m_{\rm H}) ={}&  \left\rvert\left\rvert \frac{\partial (\phi, \psi, \Sigma)}{\partial (m_{\rm L}, m_{\rm M}, m_{\rm H})} \right\rvert\right\rvert\\
    ={}& 4(m_{\rm L} m_{\rm M} + m_{\rm L} m_{\rm H} + m_{\rm M} m_{\rm H})
\end{split}
\end{align}
is the resulting reference prior for the neutrino masses.

The implicit approximation adopted here, as discussed in ref.~\cite{HeavensEvidence} is that there is translational invariance of the likelihood, that is, the likelihood does not change shape or width as a function of each parameter. While this is approximately (although certainly not exactly) true for $\Sigma$, it is interesting to consider the case for the squared mass splittings. This approximation implies that the error on $\phi$ (the small squared mass splitting) and $\psi$  (the large one) do not depend on the value of the splitting ($\phi$ or $\psi$) itself. We know that this holds around the maximum likelihood for each mass splitting: the sensitivity to the mass splitting goes roughly like $\sim \sin^2(\Delta m_{ij}^2L/E)$ where $L$ denotes the distance traveled by the neutrinos and $E$ their energy and the $L/E$ range explores more than $2\pi$ in the argument of $\sin$. However, the experimental design is very different for the large versus the small splitting. In particular, the large splitting constraints rely on accelerators and the small splitting constraints rely on nuclear reactors, which have different energies and spread physics. 

As eq.~\eqref{eq:splittings} indicates, $\phi$ is about an order of magnitude  smaller than $\psi$ (a factor  30 to be precise) and the error on $\phi$ is about an order of magnitude smaller than  $\psi$,  suggesting that experimental design might have been  tuned to reach roughly the same relative errors in each of the splittings. On the other hand, if it can be stated that the small mass splitting would have an error $\sim 10^{-6}$ eV$^2$ and the large one an error $\sim 10^{-5}$ eV$^2$ even if the  two mass spitting were of the same order of magnitude, then the likelihood is truly transationally invariant. 

It is beyond the scope of this paper to find a firm answer to this, but we believe these considerations might help the reader to appreciate the subtle but important role of priors and their relation to the physics and the mathematics underlying the issue at hand.

\providecommand{\href}[2]{#2}\begingroup\raggedright\endgroup


\begin{thebibliography}{10}

\bibitem{katrin21}
{Katrin Collaboration}, M.~{Aker}, A.~{Beglarian}, J.~{Behrens}, A.~{Berlev},
  U.~{Besserer} et~al., \emph{{Direct neutrino-mass measurement with
  sub-electronvolt sensitivity}},
  \href{https://doi.org/10.1038/s41567-021-01463-1}{\emph{Nature Physics}
  {\bfseries 18} (2022) 160}.

\bibitem{mainz}
C.~Kraus, B.~Bornschein, L.~Bornschein, J.~Bonn, B.~Flatt, A.~Kovalik et~al.,
  \emph{Final results from phase ii of the mainz neutrino mass search in
  tritium ${\beta}$ decay},
  \href{https://doi.org/10.1140/epjc/s2005-02139-7}{\emph{The European Physical
  Journal C} {\bfseries 40} (2005) 447?468}.

\bibitem{Bond:1980ha}
J.R.~Bond, G.~Efstathiou and J.~Silk, \emph{{Massive Neutrinos and the Large
  Scale Structure of the Universe}},
  \href{https://doi.org/10.1103/PhysRevLett.45.1980}{\emph{Phys. Rev. Lett.}
  {\bfseries 45} (1980) 1980}.

\bibitem{Hu:1997mj}
W.~Hu, D.J.~Eisenstein and M.~Tegmark, \emph{{Weighing neutrinos with galaxy
  surveys}}, \href{https://doi.org/10.1103/PhysRevLett.80.5255}{\emph{Phys.
  Rev. Lett.} {\bfseries 80} (1998) 5255}
  [\href{https://arxiv.org/abs/astro-ph/9712057}{{\ttfamily
  astro-ph/9712057}}].

\bibitem{1998PhRvL..80.5255H}
W.~{Hu}, D.J.~{Eisenstein} and M.~{Tegmark}, \emph{{Weighing Neutrinos with
  Galaxy Surveys}},
  \href{https://doi.org/10.1103/PhysRevLett.80.5255}{\emph{Physical Review
  Letters} {\bfseries 80} (1998) 5255}
  [\href{https://arxiv.org/abs/astro-ph/9712057}{{\ttfamily
  astro-ph/9712057}}].

\bibitem{lesgourguespastor}
J.~Lesgourgues and S.~Pastor, \emph{{Massive neutrinos and cosmology}},
  \href{https://doi.org/10.1016/j.physrep.2006.04.001}{\emph{Phys.Rept.}
  {\bfseries 429} (2006) 307}
  [\href{https://arxiv.org/abs/astro-ph/0603494}{{\ttfamily
  astro-ph/0603494}}].

\bibitem{JimenezKitching}
R.~{Jimenez}, T.~{Kitching}, C.~{Pe{\~n}a-Garay} and L.~{Verde}, \emph{{Can we
  measure the neutrino mass hierarchy in the sky?}},
  \href{https://doi.org/10.1088/1475-7516/2010/05/035}{\emph{JCAP} {\bfseries
  2010} (2010) 035} [\href{https://arxiv.org/abs/1003.5918}{{\ttfamily
  1003.5918}}].

\bibitem{Wagner:2012sw}
C.~Wagner, L.~Verde and R.~Jimenez, \emph{{Effects of the neutrino mass
  splitting on the non-linear matter power spectrum}},
  \href{https://doi.org/10.1088/2041-8205/752/2/L31}{\emph{Astrophys. J.}
  {\bfseries 752} (2012) L31}
  [\href{https://arxiv.org/abs/1203.5342}{{\ttfamily 1203.5342}}].

\bibitem{eBOSSresults}
S.~Alam, M.~Aubert, S.~Avila, C.~Balland, J.E.~Bautista, M.A.~Bershady et~al.,
  \emph{Completed sdss-iv extended baryon oscillation spectroscopic survey:
  Cosmological implications from two decades of spectroscopic surveys at the
  apache point observatory},
  \href{https://doi.org/10.1103/physrevd.103.083533}{\emph{Physical Review D}
  {\bfseries 103} (2021) }.

\bibitem{1911.09073}
N.~Palanque-Delabrouille, C.~Yeche, N.~Schoneberg, J.~Lesgourgues,
  M.~Walther, S.~Chabanier et~al., \emph{Hints, neutrino bounds, and wdm
  constraints from sdss dr14 lyman-$\alpha$ and planck full-survey data},
  \href{https://doi.org/10.1088/1475-7516/2020/04/038}{\emph{Journal of
  Cosmology and Astroparticle Physics} {\bfseries 2020} (2020) 038?038}.

\bibitem{Fergus}
F.~{Simpson}, R.~{Jimenez}, C.~{Pena-Garay} and L.~{Verde}, \emph{{Strong
  Bayesian evidence for the normal neutrino hierarchy}},
  \href{https://doi.org/10.1088/1475-7516/2017/06/029}{\emph{JCAP} {\bfseries
  2017} (2017) 029} [\href{https://arxiv.org/abs/1703.03425}{{\ttfamily
  1703.03425}}].

\bibitem{Murayama:2003ci}
H.~Murayama and C.~Pena-Garay, \emph{{Neutrinoless double beta decay in light
  of SNO salt data}},
  \href{https://doi.org/10.1103/PhysRevD.69.031301}{\emph{Phys. Rev. D}
  {\bfseries 69} (2004) 031301}
  [\href{https://arxiv.org/abs/hep-ph/0309114}{{\ttfamily hep-ph/0309114}}].

\bibitem{Giuliani:2019uno}
{\scshape APPEC Committee} collaboration, \emph{{Double Beta Decay APPEC
  Committee Report}},  \href{https://arxiv.org/abs/1910.04688}{{\ttfamily
  1910.04688}}.

\bibitem{Cox}
R.T.~Cox, \emph{Probability, frequency and reasonable expectation},
  {\emph{{Am.J.Th.Phys}} {\bfseries 14} (1946) 1}.

\bibitem{Jaffe:1996}
A.~{Jaffe}, \emph{{$H_0$ and Odds on Cosmology}},
  \href{https://doi.org/10.1086/177950}{\emph{Astrophys.J.} {\bfseries 471}
  (1996) 24} [\href{https://arxiv.org/abs/arXiv:astro-ph/9501070}{{\ttfamily
  arXiv:astro-ph/9501070}}].

\bibitem{Trotta2}
R.~{Trotta}, \emph{{Applications of Bayesian model selection to cosmological
  parameters}},
  \href{https://doi.org/10.1111/j.1365-2966.2007.11738.x}{\emph{Mon.Not.R.Astron.Soc}
  {\bfseries 378} (2007) 72}
  [\href{https://arxiv.org/abs/arXiv:astro-ph/0504022}{{\ttfamily
  arXiv:astro-ph/0504022}}].

\bibitem{LiddleEvidence04}
A.R.~{Liddle}, \emph{{How many cosmological parameters?}},
  \href{https://doi.org/10.1111/j.1365-2966.2004.08033.x}{\emph{Mon.Not.R.Astron.Soc}
  {\bfseries 351} (2004) L49}
  [\href{https://arxiv.org/abs/arXiv:astro-ph/0401198}{{\ttfamily
  arXiv:astro-ph/0401198}}].

\bibitem{mackay2003information}
D.J.~MacKay, \emph{Information theory, inference and learning algorithms},
  Cambridge university press (2003).

\bibitem{objectivebayesian}
A.F.~{Heavens} and E.~{Sellentin}, \emph{{Objective Bayesian analysis of
  neutrino masses and hierarchy}},
  \href{https://doi.org/10.1088/1475-7516/2018/04/047}{\emph{JCAP} {\bfseries
  2018} (2018) 047} [\href{https://arxiv.org/abs/1802.09450}{{\ttfamily
  1802.09450}}].

\bibitem{referee_request}
M.~{Gerbino}, M.~{Lattanzi}, O.~{Mena} and K.~{Freese}, \emph{{A novel approach
  to quantifying the sensitivity of current and future cosmological datasets to
  the neutrino mass ordering through Bayesian hierarchical modeling}},
  \href{https://doi.org/10.1016/j.physletb.2017.10.052}{\emph{Physics Letters
  B} {\bfseries 775} (2017) 239}
  [\href{https://arxiv.org/abs/1611.07847}{{\ttfamily 1611.07847}}].

\bibitem{Blennow:2013kga}
M.~Blennow, \emph{{On the Bayesian approach to neutrino mass ordering}},
  \href{https://doi.org/10.1007/JHEP01(2014)139}{\emph{JHEP} {\bfseries 01}
  (2014) 139} [\href{https://arxiv.org/abs/1311.3183}{{\ttfamily 1311.3183}}].

\bibitem{HannestadSchwetz}
S.~{Hannestad} and T.~{Schwetz}, \emph{{Cosmology and the neutrino mass
  ordering}}, \href{https://doi.org/10.1088/1475-7516/2016/11/035}{\emph{JCAP}
  {\bfseries 2016} (2016) 035}
  [\href{https://arxiv.org/abs/1606.04691}{{\ttfamily 1606.04691}}].

\bibitem{Vagnozzi17}
S.~{Vagnozzi}, E.~{Giusarma}, O.~{Mena}, K.~{Freese}, M.~{Gerbino}, S.~{Ho}
  et~al., \emph{{Unveiling {\ensuremath{\nu}} secrets with cosmological data:
  Neutrino masses and mass hierarchy}},
  \href{https://doi.org/10.1103/PhysRevD.96.123503}{\emph{PRD} {\bfseries 96}
  (2017) 123503} [\href{https://arxiv.org/abs/1701.08172}{{\ttfamily
  1701.08172}}].

\bibitem{LongRaveri}
A.J.~{Long}, M.~{Raveri}, W.~{Hu} and S.~{Dodelson}, \emph{{Neutrino mass
  priors for cosmology from random matrices}},
  \href{https://doi.org/10.1103/PhysRevD.97.043510}{\emph{PRD} {\bfseries 97}
  (2018) 043510} [\href{https://arxiv.org/abs/1711.08434}{{\ttfamily
  1711.08434}}].

\bibitem{ChoudhuryHannestad20}
S.R.~{Choudhury} and S.~{Hannestad}, \emph{{Updated results on neutrino mass
  and mass hierarchy from cosmology with Planck 2018 likelihoods}},
  \href{https://doi.org/10.1088/1475-7516/2020/07/037}{\emph{JCAP} {\bfseries
  2020} (2020) 037} [\href{https://arxiv.org/abs/1907.12598}{{\ttfamily
  1907.12598}}].

\bibitem{GariazzoMena19}
S.~{Gariazzo} and O.~{Mena}, \emph{{Cosmology-marginalized approaches in
  Bayesian model comparison: The neutrino mass as a case study}},
  \href{https://doi.org/10.1103/PhysRevD.99.021301}{\emph{PRD} {\bfseries 99}
  (2019) 021301} [\href{https://arxiv.org/abs/1812.05449}{{\ttfamily
  1812.05449}}].

\bibitem{Hergt}
L.T.~{Hergt}, W.J.~{Handley}, M.P.~{Hobson} and A.N.~{Lasenby}, \emph{{Bayesian
  evidence for the tensor-to-scalar ratio r and neutrino masses
  m$_{{\ensuremath{\nu}}}$ : Effects of uniform versus logarithmic priors}},
  \href{https://doi.org/10.1103/PhysRevD.103.123511}{\emph{PRD} {\bfseries 103}
  (2021) 123511} [\href{https://arxiv.org/abs/2102.11511}{{\ttfamily
  2102.11511}}].

\bibitem{Mahony}
C.~{Mahony}, B.~{Leistedt}, H.V.~{Peiris}, J.~{Braden}, B.~{Joachimi},
  A.~{Korn} et~al., \emph{{Target neutrino mass precision for determining the
  neutrino hierarchy}},
  \href{https://doi.org/10.1103/PhysRevD.101.083513}{\emph{PRD} {\bfseries 101}
  (2020) 083513} [\href{https://arxiv.org/abs/1907.04331}{{\ttfamily
  1907.04331}}].

\bibitem{HeavensEvidence}
A.F.~{Heavens}, T.D.~{Kitching} and L.~{Verde}, \emph{{On model selection
  forecasting, dark energy and modified gravity}},
  \href{https://doi.org/10.1111/j.1365-2966.2007.12134.x}{\emph{Mon.Not.R.Astron.Soc.}
  {\bfseries 380} (2007) 1029}
  [\href{https://arxiv.org/abs/arXiv:astro-ph/0703191}{{\ttfamily
  arXiv:astro-ph/0703191}}].

\bibitem{gonzalez-garciaglobal21}
M.C.~{Gonzalez-Garcia}, M.~{Maltoni} and T.~{Schwetz}, \emph{{NuFIT:
  Three-Flavour Global Analyses of Neutrino Oscillation Experiments}},
  {\emph{arXiv e-prints} (2021) arXiv:2111.03086}
  [\href{https://arxiv.org/abs/2111.03086}{{\ttfamily 2111.03086}}].

\bibitem{katrin2019}
M.~Aker, K.~Altenmuller, M.~Arenz, M.~Babutzka, J.~Barrett, S.~Bauer et~al.,
  \emph{Improved upper limit on the neutrino mass from a direct kinematic
  method by katrin},
  \href{https://doi.org/10.1103/physrevlett.123.221802}{\emph{Physical Review
  Letters} {\bfseries 123} (2019) }.

\bibitem{Elgaroy:2002bi}
O.~Elgaroy et~al., \emph{{A New limit on the total neutrino mass from the 2dF
  galaxy redshift survey}},
  \href{https://doi.org/10.1103/PhysRevLett.89.061301}{\emph{Phys. Rev. Lett.}
  {\bfseries 89} (2002) 061301}
  [\href{https://arxiv.org/abs/astro-ph/0204152}{{\ttfamily
  astro-ph/0204152}}].

\bibitem{WMAP:2012nax}
{\scshape WMAP} collaboration, \emph{{Nine-Year Wilkinson Microwave Anisotropy
  Probe (WMAP) Observations: Cosmological Parameter Results}},
  \href{https://doi.org/10.1088/0067-0049/208/2/19}{\emph{Astrophys. J. Suppl.}
  {\bfseries 208} (2013) 19} [\href{https://arxiv.org/abs/1212.5226}{{\ttfamily
  1212.5226}}].

\bibitem{Planck2013cosmo}
{Planck Collaboration}, P.A.R.~{Ade}, N.~{Aghanim}, C.~{Armitage-Caplan},
  M.~{Arnaud}, M.~{Ashdown} et~al., \emph{{Planck 2013 results. XVI.
  Cosmological parameters}},
  \href{https://doi.org/10.1051/0004-6361/201321591}{\emph{A\&A} {\bfseries
  571} (2014) A16} [\href{https://arxiv.org/abs/1303.5076}{{\ttfamily
  1303.5076}}].

\bibitem{Planck2015cosmo}
{Planck Collaboration}, P.A.R.~{Ade}, N.~{Aghanim}, M.~{Arnaud}, M.~{Ashdown},
  J.~{Aumont} et~al., \emph{{Planck 2015 results. XIII. Cosmological
  parameters}}, \href{https://doi.org/10.1051/0004-6361/201525830}{\emph{A\&A}
  {\bfseries 594} (2016) A13}
  [\href{https://arxiv.org/abs/1502.01589}{{\ttfamily 1502.01589}}].

\bibitem{2016Cuesta}
A.J.~{Cuesta}, V.~{Niro} and L.~{Verde}, \emph{{Neutrino mass limits: Robust
  information from the power spectrum of galaxy surveys}},
  \href{https://doi.org/10.1016/j.dark.2016.04.005}{\emph{Physics of the Dark
  Universe} {\bfseries 13} (2016) 77}
  [\href{https://arxiv.org/abs/1511.05983}{{\ttfamily 1511.05983}}].

\bibitem{KR95}
R.E.~Kass and A.E.~Raftery, \emph{Bayes factors}, {\emph{Journal of the
  American Statistical Association} {\bfseries 90} (1995) 773}.

\bibitem{Bernardo}
J.M.~Bernardo, \emph{The concept of exchangeability and its applications},
  {\emph{Far East J. Mathematical Sciences} {\bfseries 4} (2006) 111}.

\bibitem{Finetti31}
B.~de~Finetti, \emph{Funcione caratteristica di un fenomeno aleatorio.},
  {\emph{Atti Dela Reale Accademia Nazionale Dei Lincei, serie 6} {\bfseries 4}
  (1931) 251}.

\bibitem{BernardoSmith09}
J.M.~Bernardo and A.F.~Smith, \emph{Bayesian theory}, {\emph{John Wiley \&
  Sons} {\bfseries 405} (2009) }.

\bibitem{reference_prior}
J.O.~Berger, J.M.~Bernardo and D.~Sun, \emph{The formal definition of reference
  priors}, \href{https://doi.org/10.1214/07-aos587}{\emph{The Annals of
  Statistics} {\bfseries 37} (2009) }.

\end{thebibliography}

\end{document}